\documentclass[aps,prb,twocolumn,superscriptaddress,floatfix]{revtex4-2}

\usepackage{amsmath,amssymb}
\usepackage{graphicx}
\usepackage{booktabs}
\usepackage{siunitx}
\usepackage{hyperref}
\usepackage{xcolor}
\usepackage{multirow}

\graphicspath{{figs/}}

\newcommand{\ehull}{E_\mathrm{hull}}

\newcommand{\muBA}{\mu_\mathrm{B}/\text{\AA}^3}

\begin{document}

\title{High-throughput Discovery of Magnetic Rare Earth Transition Metal Alloys}

\author{Shuo Tao}
\email{stao@charlotte.edu}
\affiliation{Department of Mechanical Engineering and Engineering Science, University of North Carolina at Charlotte, Charlotte, NC 28223, USA}

\author{Osman Goni Ridwan}
\affiliation{Department of Mechanical Engineering and Engineering Science, University of North Carolina at Charlotte, Charlotte, NC 28223, USA}

\author{Liqin Ke}
\affiliation{Department of Materials Science and Engineering, University of Virginia, Charlottesville, VA 22904, USA}

\author{Qiang Zhu}
\email{qzhu8@charlotte.edu}
\affiliation{Department of Mechanical Engineering and Engineering Science, University of North Carolina at Charlotte, Charlotte, NC 28223, USA}
\affiliation{North Carolina Battery Complexity, Autonomous Vehicle and Electrification (BATT CAVE) Research Center, Charlotte, NC 28223, USA}

\date{\today}

\begin{abstract}
We present an accelerated materials discovery framework that combines diffusion-based crystal structure generation with hierarchical screening to identify new rare-earth--transition-metal magnets simultaneously achieving high magnetization and thermodynamic stability. Using this workflow, we systematically explored over 3000 binary (R-T) and ternary (R-T-T$'$) compositions spanning R~$\in \{\text{Y, Sm}\}$, T~$\in \{\text{Fe, Co, Ni}\}$, and T$' \in \{\text{Ti, V, Cr, Mn, Cu, Zn}\}$, and filtered approximately 240{,}000 generated crystal structures through machine-learning interatomic potential prescreening and spin-polarized density functional theory validation. We identify 300+ low-energy magnetic candidates within 0.1~eV/atom above the convex hull at the DFT level, including 5 thermodynamically stable phases. The highest saturation magnetization reaches ${\sim}1.8$~T in Fe-rich binary and ternary phases (SmFe$_{12}$, YFe$_{12}$, YFe$_{18}$Ti and Sm$_2$Fe$_{16}$Mn). Symmetry analysis reveals that the majority of ternary candidates are subgroup derivatives of known binary prototypes through Wyckoff site splitting that accommodates T$'$ substitution. Site-resolved magnetic moment analysis further shows that Mn aligns ferromagnetically with the Fe sublattice with minimal magnetization loss, whereas Cr couples antiferromagnetically, providing systematic guidance for dopant selection. These findings demonstrate a generalizable strategy for targeted magnetic materials discovery and suggest that extending generative searches to larger unit cells ($>$20 atoms) with higher Fe fractions is a promising route toward stable phases with saturation magnetization exceeding 1.8~T.
\end{abstract}

\maketitle

\section{Introduction}\label{sec:intro}

Rare Earth Transition Metal (RE-TM) intermetallic compounds form the
backbone of modern permanent-magnet technology.
The discovery of SmCo$_5$ in 1967 established the first high-performance RE magnet family,
exploiting the large magnetocrystalline anisotropy arising from the
Sm's 4$f$ electrons in the hexagonal CaCu$_5$-type structure~\cite{strnatFamilyNewCobaltbase1967}.
The subsequent development of Nd$_2$Fe$_{14}$B in 1984  achieved record energy products by replacing
expensive Co with abundant Fe, enabled by the addition of
interstitial B to stabilize a favorable tetragonal structure~\cite{sagawaNewMaterialPermanent1984}.
Together with the Sm$_2$Co$_{17}$ family, these materials now
dominate applications ranging from electric-vehicle traction motors
and wind-turbine generators to hard-disk drives and medical devices
\cite{gutfleischMagneticMaterialsDevices2011, coeyPerspectiveProspectsRare2020}.

Despite their technological maturity, the known RE-TM permanent magnet phases represent only a small fraction of the accessible
composition--structure space~\cite{snyder2026possibility}.
Experimental discovery has historically proceeded by intuition-guided synthesis, an approach that is slow, costly, and inherently biased toward well-explored prototype families such as
CaCu$_5$ (1:5), Th$_2$Zn$_{17}$ (2:17),
and ThMn$_{12}$ (1:12) types \cite{coeyPermanentMagnetism1997}.
Meanwhile, supply-chain vulnerabilities have intensified the search for new phases
with reduced rare-earth content or improved properties at comparable
compositions~\cite{skokovHeavyRareEarth2018}.

High-throughput density functional theory (DFT) screening,
combined with large-scale materials databases, has emerged as a
powerful strategy for accelerating materials discovery
\cite{curtaroloHighthroughputHighwayComputational2013, hautierFindingNeedleHaystack2019}.
The Materials Project \cite{jainCommentaryMaterialsProject2013} and the AFLOW consortium
now provide computed properties for hundreds of thousands of
inorganic phases, including magnetic ground states
\cite{hortonHighthroughputPredictionGroundstate2019}.
Targeted DFT campaigns have screened specific prototype families
for permanent-magnet candidates:
Sanvito et al.\ screened over 236{,}000 Heusler prototypes and
identified 20 stable magnetic intermetallics, subsequently
synthesizing two new magnets including Co$_2$MnTi
\cite{sanvitoAcceleratedDiscoveryNew2017};
K\"{o}rner et al.\ evaluated 1{,}280 ThMn$_{12}$-type phases
for rare-earth-lean hard magnets with promising
Ce-based candidates \cite{kornerTheoreticalScreeningIntermetallic2016};
and Drebov et al.\ developed an HT methodology for
RE-TM compounds covering the 1:5, 2:14:1, and 2:17 families
\cite{drebovInitioScreeningMethodology2013}.
However, these prototype-based searches are inherently limited
to known crystal structure types and cannot discover phases with
entirely new topologies.

Machine learning (ML) has further accelerated computational
magnetic-materials design, particularly for predicting the
Curie temperature ($T_\mathrm{C}$) and magnetic ordering.
Nelson and Sanvito trained composition-based models on
${\sim}2{,}500$ experimental $T_\mathrm{C}$ values,
achieving prediction accuracy of ${\sim}50$~K
\cite{nelsonPredictingCurieTemperature2019}.
Long et al.\ classified ferromagnetic versus antiferromagnetic
compounds and predicted $T_\mathrm{C}$ using random forests
\cite{longAcceleratingApproachDesigning2021}.
Physics-informed ML models have been attempted to capture electronic-structure
trends underlying $T_\mathrm{C}$ in rare-earth intermetallics
\cite{singhPhysicsinformedMachinelearningPrediction2023,
nguyenRegressionbasedModelEvaluation2019},
while the Northeast Materials Database (NEMAD) demonstrated
large-language-model-assisted extraction of experimental
magnetic data for 67{,}573 entries \cite{itaniNortheastMaterialsDatabase2025}.
Beyond property prediction, ML has been integrated into
structure-search workflows for discovering rare-earth-free magnets.
Xia et al.\ combined crystal graph convolutional neural networks
(CGCNN) with adaptive genetic algorithms and DFT to discover
Fe$_3$CoB$_2$ with high magnetic anisotropy and saturation
polarization \cite{xiaAcceleratingDiscoveryNovel2022}, and subsequently
extended this framework to identify promising compounds in the
Fe-Co-P \cite{liaoMachineLearningacceleratedDiscovery2024},
Fe-Co-C \cite{xiaMachineLearningAssisted2024},
Fe-Co-Zr \cite{xiaAcceleratedDiscoveryDesign2026},
and Ce-Co-Cu \cite{xiaSearchStableLowenergy2025} systems.
In these approaches, ML serves as a surrogate for formation-energy
evaluation, enabling rapid prescreening of substitutional structures
generated from known prototypes.

More recently, another emerging direction is the use of generative deep-learning models to propose entirely new crystal structures, bypassing the
reliance on existing prototype databases altogether.
\texttt{MatterGen}, a diffusion-based generative model trained on all known stable inorganic materials, can generate novel structures conditioned
on target chemistry, symmetry, or property constraints such as
magnetization \cite{zeniGenerativeModelInorganic2025}.
Coupled with universal machine-learning interatomic potentials such
as \texttt{MatterSim} \cite{yangMatterSimDeepLearning2024}, which provide near-DFT accuracy at a fraction of the computational cost, generative structure
prediction can in principle explore regions of configuration space
inaccessible to prototype-based enumeration.

Despite these exciting developments, a systematic generative exploration of the RE-TM magnetic alloy space remains largely unexplored, as the interplay of $4f$-$3d$ exchange coupling, competing magnetic sublattices, and the sensitivity of phase stability to rare-earth $4f$ electron treatment pose additional challenges. In this work, we address this gap by deploying a four-stage
high-throughput pipeline 
to systematically explore TM-rich binary (R-T) and ternary (R-T-T$'$)
compositions across R~$= \{\text{Y}, \text{Sm}\}$,
T~$= \{\text{Fe}, \text{Co}, \text{Ni}\}$,
and T$' = \{\text{Ti}, \text{V}, \text{Cr}, \text{Mn}, \text{Cu}, \text{Zn}\}$.
Starting from approximately 240{,}000 generated structures spanning
3{,}138 stoichiometries, we identify 344 low-energy magnetic candidates.
We analyze structural trends, Wyckoff-site substitution patterns,
and magnetization properties across the candidate set, concluded by the remarks on further method development towards achieving new powerful permanent magnets.

\section{Methods}\label{sec:methods}

Our high-throughput screening pipeline consists of four stages:
(i) systematic enumeration of RE-TM compositions that favor high magnetization,
(ii) rapid crystal structure sampling with \texttt{MatterGen},
(iii) stability prescreening with \texttt{MatterSim},
and (iv) spin-polarized DFT validation.

\subsection{Chemical Space Definition}\label{sec:chemspace}

To discover novel high-performance permanent magnets, we highlight three main design principles. First, we target transition-metal-rich RE--TM compositions where the saturation magnetization ($M_\mathrm{s}$) is governed by a dense, exchange-coupled $3d$ sublattice of T~$\in \{\text{Fe, Co, Ni}\}$. Second, rather than sampling the entire rare-earth series, we select two representative elements, $\mathrm{R} \in \{\text{Y}, \text{Sm}\}$. Yttrium serves as an $f^0$ non-magnetic baseline reference to isolate and evaluate the intrinsic contribution of the $3d$ transition-metal sublattice to magnetic anisotropy and magnetization. Samarium represents the $4f$ rare-earth archetype, renowned for providing strong uniaxial magnetocrystalline anisotropy through strong $4f\text{--}3d$ exchange coupling in benchmark permanent-magnet phases (e.g., $\text{SmCo}_5$ and $\text{Sm}_2\text{Fe}_{17}$). Third, we optionally utilize a set of $3d$ transition metal (Ti, V, Cr, Mn, Cu, Zn) as the additives to stabilize metastable high-anisotropy crystal phases.

For binary R--T compositions, we enforce a transition-metal-rich threshold of
$R_\mathrm{TM}=N_\mathrm{T} / (N_\mathrm{R} + N_\mathrm{T}) \geq 0.75$ with a unit cell constraint of 
$N_\mathrm{R} + N_\mathrm{T} \leq 20$ atoms at most. 
For ternary R--T--$\mathrm{T}'$ compositions, we additionally enforce
$R_\mathrm{TM}=N_\mathrm{T} / (N_\mathrm{R} + N_\mathrm{T} + N_{\mathrm{T}'}) \geq 0.80$
to ensure that the primary magnetic element dominates the $3d$ sublattice, while maintaining a total primitive cell size of
$N_\mathrm{R} + N_\mathrm{T} + N_{\mathrm{T}'} \leq 20$.
All stoichiometries are reduced to their primitive form via greatest common divisor normalization.
This systematic enumeration yields 186 unique binary and 2{,}952 unique ternary compositions as summarized in Table~\ref{tab:chemspace}.

\begin{table}[htbp]
\caption{Chemical space summary.}
\label{tab:chemspace}
\begin{ruledtabular}
\begin{tabular}{lcc}
 & Binary (R-T) & Ternary (R-T-T$'$) \\
\midrule
Max atoms/cell          & 20    & 20 \\
$R_\mathrm{TM}$ cutoff & $\geq 0.75$ & $\geq 0.75$ \\
T dominance             & ---   & $N_\mathrm{T}/(N_\mathrm{T}+N_{\mathrm{T}'})\geq 0.80$ \\
Total compositions      & 186   & 2{,}952 \\
\end{tabular}
\end{ruledtabular}
\end{table}

\begin{table*}[htbp]
\caption{High-throughput screening funnel.
``Generated'' counts all \texttt{MatterGen} generated structures;
``unique'' counts after duplicate removal;
``passed MLP'' counts structures below the $\ehull^\mathrm{MLP\text{-}ref}$ threshold
(\SI{0.1}{eV/atom} for binary, \SI{0.06}{eV/atom} for ternary);
``final candidates'' counts structures with $\ehull^\mathrm{DFT\text{-}ref} < \SI{0.1}{eV/atom}$
and space-group number $\geq 75$.
``On hull'' denotes structures with $\ehull^\mathrm{DFT} = 0$
(thermodynamically stable on the true DFT hull).}
\label{tab:funnel}
\begin{ruledtabular}
\begin{tabular}{llcrrrrr}
RE & System & MLP threshold & Generated & Passed MLP & DFT computed & Final candidates & On hull \\
\midrule
Y    & Binary  & 0.1  &  16{,}032 & 2{,}849 & 2{,}841 &  60 & 1 \\
Y    & Ternary & 0.06 &  74{,}938 & 1{,}581 & 1{,}576 & 125 & 3 \\
Sm   & Binary  & 0.1  &  16{,}032 & 1{,}078 & 1{,}078 &  50 & 0 \\
Sm   & Ternary & 0.06 & 132{,}768 & 1{,}670 & 1{,}667 & 109 & 1 \\
\midrule
\multicolumn{3}{l}{Total} & 239{,}770 & 7{,}178 & 7{,}162 & 344 & 5 \\
\end{tabular}
\end{ruledtabular}
\end{table*}

\subsection{Crystal Structure Sampling}\label{sec:csp}

Next, crystal structures for each composition were generated using \texttt{MatterGen}. To explore different cell sizes within the 20-atom training limit,
each reduced formula was expanded to all valid supercell multiples. For example, a 6-atom primitive cell (e.g., RT$_5$) was generated at 6, 12, and 18~atoms/cell; a 9-atom cell (e.g., R$_2$T$_7$) at 9 and
18~atoms/cell; and a 19-atom cell (e.g., R$_2$T$_{17}$) at 19~atoms/cell only.
The number of candidate structures generated for each composition was set
proportional to the total atom count across all its supercells, at a rate of
6.0~structures per atom.
As shown in Table~\ref{tab:funnel}, this procedure yielded approximately $4.7 \times 10^5$ candidate structures
in total across all compositions and both RE batches.

\subsection{Stability Prescreening}\label{sec:prescreen}


All \texttt{MatterGen} generated structures were prescreened using \texttt{MatterSim-v1.0.0-5M},
a machine-learning interatomic potential (MLIP) based on the M3GNet architecture~\cite{chen2022universal},
trained on the Materials Project DFT-GGA dataset.
Each candidate structure was relaxed with the corresponding MatterSim potential using force convergence threshold \SI{0.01}{eV/\angstrom}, a maximum of 500 ionic steps, and the
ExpCellFilter for simultaneous cell and position optimization.
Duplicate structures within each composition were removed using pymatgen's
\texttt{StructureMatcher}
(length tolerance 0.2, site tolerance 0.2, angle tolerance \SI{5}{\degree})~\cite{pymatgen-2013}.
The energy above the MP-referenced convex hull, $\ehull^\mathrm{MLP\text{-}ref}$,
was computed against MP GGA reference energies,
and structures with $\ehull^\mathrm{MLP\text{-}ref}$ below a threshold
(\SI{0.1}{eV/atom} for binary, \SI{0.06}{eV/atom} for ternary systems)
were retained for DFT validation.
Table~\ref{tab:funnel} summarizes the prescreening outcomes.

\subsection{DFT validation}\label{sec:dft}

All structures that passed prescreening were relaxed with
spin-polarized DFT using \texttt{VASP}~\cite{Vasp-PRB-1996} and PAW-PBE pseudopotentials~\cite{PBE-PRL-1996}.
Before VASP input generation, each structure was symmetrized using \texttt{PyXtal}~\cite{pyxtal}
with progressive symmetry tolerances ($5\times10^{-2}$ to $1\times10^{-5}$).
Spin polarization (\texttt{ISPIN}~=~2) was enabled with initial magnetic moments
following a ferrimagnetic RE--TM convention:
rare-earth moments were initialized antiparallel to the TM sublattice
(e.g., Sm: $-5.0\,\mu_\mathrm{B}$; Fe: $+2.2\,\mu_\mathrm{B}$;
Co: $+2.0\,\mu_\mathrm{B}$; Ni: $+2.0\,\mu_\mathrm{B}$),
reflecting the expected ferrimagnetic coupling in RE--TM intermetallics.

\begin{figure}[htbp]
\centering
\includegraphics[width=\columnwidth]{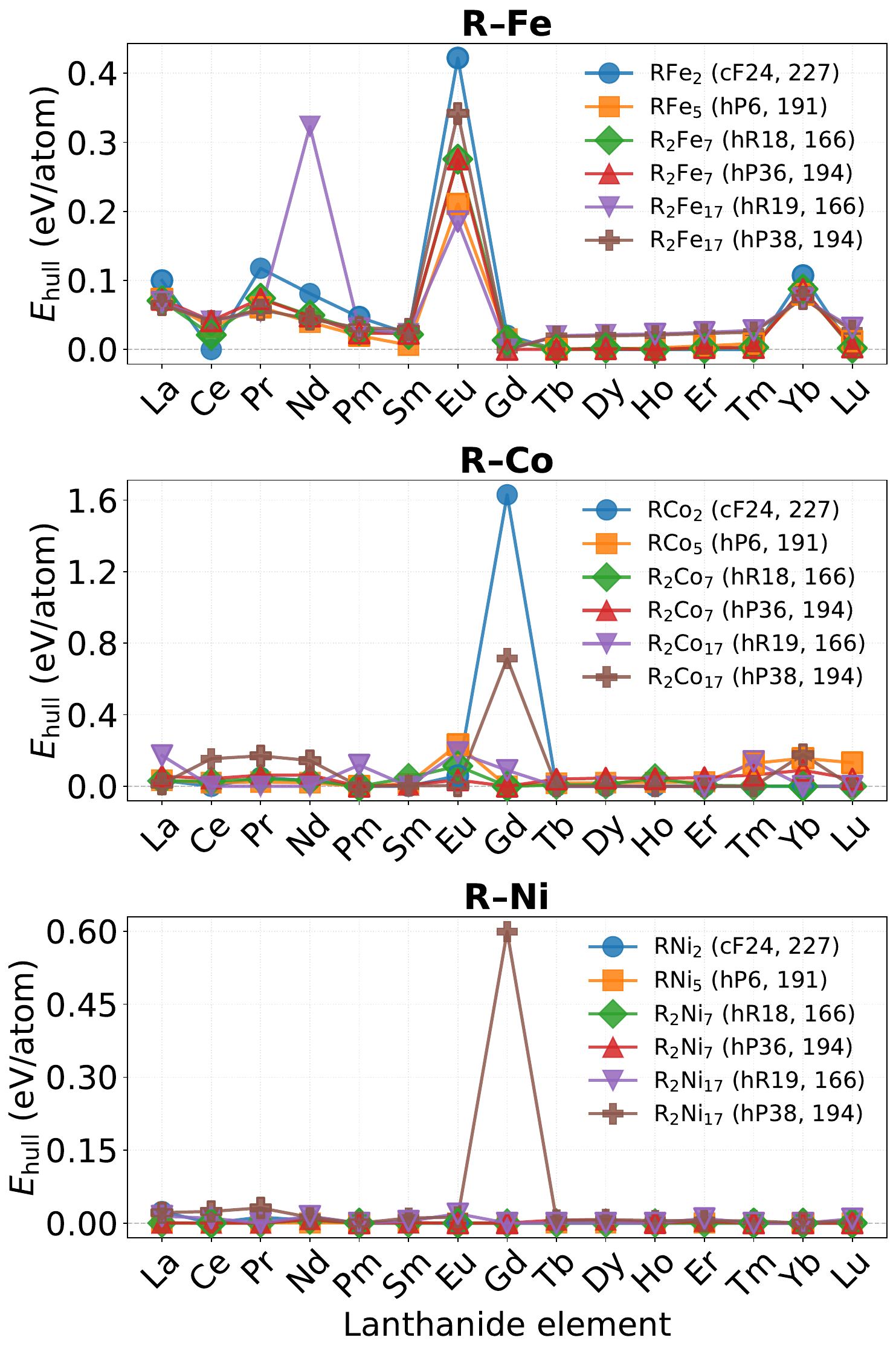}
\caption{Materials Project $\ehull$ values for binary R--TM phases across the lanthanide series (La--Lu), plotted for the R--Fe (top),
R--Co (middle), and R--Ni (bottom) systems. Each prototype is labeled by
its reduced chemical formula, Pearson symbol, and space-group number.}
\label{fig:mp_validate}
\end{figure}

The choice of RE pseudopotentials requires particular care. For Y, which has no $4f$ electrons, the standard PAW-PBE potential (11 valence electrons) is used directly. For lanthanides, however, the treatment of the localized $4f$ shell introduces systematic errors in the Materials Project (MP) reference energies. To quantify this, we computed the MP $\ehull$ for all known binary R--TM prototypes across the full lanthanide series (Fig.~\ref{fig:mp_validate}). The validation reveals anomalously large $\ehull$ values for several experimentally confirmed stable or low-energy phases, most prominently in Gd systems
(e.g., GdCo$_2$: $\ehull = 1.63$~eV/atom;
Gd$_2$Co$_{17}$: 0.71~eV/atom;
Gd$_2$Ni$_{17}$: 0.60~eV/atom)
and to a lesser extent in Eu--Fe compounds.
These errors originate from the GGA treatment of the half-filled Gd~$4f^7$ shell and render the MP convex hull unreliable for thermodynamic screening of these elements.

Among all lanthanides, Gd was our initial element of interest owing to its large $4f$ moment. Given the anomalous MP reference energies described above, we instead selected Sm---the nearest lanthanide neighbor to the left of Gd whose MP reference phases remain consistent across all three TM series (Fig.~\ref{fig:mp_validate}). We did not select Tb, the nearest neighbor to the right of Gd, for reasons related to the $4f$-electron treatment detailed below.

The $4f$ electrons in lanthanides are highly localized and contribute primarily to the on-site RE magnetic moment rather than to chemical bonding, which is governed by RE~$5d/6s$ and TM~$3d$ hybridization. Treating the $4f$ states explicitly as valence electrons without a Hubbard~$U$ correction leads to spurious pinning of the $4f$ levels near the Fermi energy and non-integer $f$-occupations during the self-consistent cycle. Applying DFT+$U$ resolves this issue but, in turn, requires spin--orbit coupling (SOC) to capture the orbital angular momentum contribution to the total $4f$ moment. Neither DFT+$U$ nor SOC is employed in the MP reference calculations for RE--TM systems---MP uses $4f$-in-valence PAW potentials only for Ce and Gd among all lanthanides. We therefore adopt the trivalent Sm\_3 PAW potential ($4f$-frozen, 11 valence electrons), which is the pymatgen \texttt{MPRelaxSet} default for Sm, ensuring full energy consistency with the MP reference convex hull.

Beyond thermodynamic consistency, the $f$-frozen Sm\_3 potential also yields physically meaningful magnetization values. By Hund's rules, Sm$^{3+}$ has configuration $4f^5$ with $S = 5/2$, $L = 5$, and $J = |L - S| = 5/2$ (less than half-filled); the spin and orbital angular momenta couple antiparallel, resulting in a small net $4f$ moment ($g_J J \approx 0.71\,\mu_\mathrm{B}$ for the free ion). The dominant magnetization in Sm--TM compounds therefore arises from the $3d$ sublattice and the $5d$--$3d$ exchange interaction, both of which are fully captured by the $f$-frozen potential. This stands in contrast to Tb ($4f^8$, $S = 3$, $L = 3$, $J = L + S = 6$; more than half-filled), where spin and orbital moments are parallel, giving a large net $4f$ moment ($g_J J \approx 9.0\,\mu_\mathrm{B}$) that would be entirely absent in a frozen-core treatment. For this reason, we restrict the present analysis to Y and Sm, where both formation energies and magnetization values are reliable within the chosen POTCAR framework.

Finally, the DFT energy above the MP-referenced convex hull, $\ehull^\mathrm{DFT\text{-}ref}$,
was computed for each relaxed structure against MP reference phases.
Candidates were selected by requiring
$\ehull^\mathrm{DFT\text{-}ref} < \SI{0.1}{eV/atom}$
and space-group number $\ge 75$
(retaining tetragonal, trigonal, hexagonal, and cubic symmetries).
A second quantity, the true DFT energy above hull $\ehull^\mathrm{DFT}$,
was computed by constructing the convex hull from both
MP reference phases and all generated structures;
this quantity is non-negative by construction and identifies
thermodynamically stable candidates ($\ehull^\mathrm{DFT} = 0$).
Each candidate was classified against the AFLOW prototype library~\cite{mehl2017aflow}: binary structures were matched directly,
while ternary structures were matched by merging the T and T$'$ sublattices
into a single TM site (``binary derived'').
This classification reveals the parent binary prototype from which each ternary structure derives
and identifies which Wyckoff site the dopant T$'$ occupies.
Total magnetic moment and cell volume were extracted from the VASP calculations,
and the magnetization $M/V$ ($\muBA$) was computed.

\section{Results}\label{sec:results}

\begin{figure*}[htbp]
  \centering
  \includegraphics[width=0.95\textwidth]{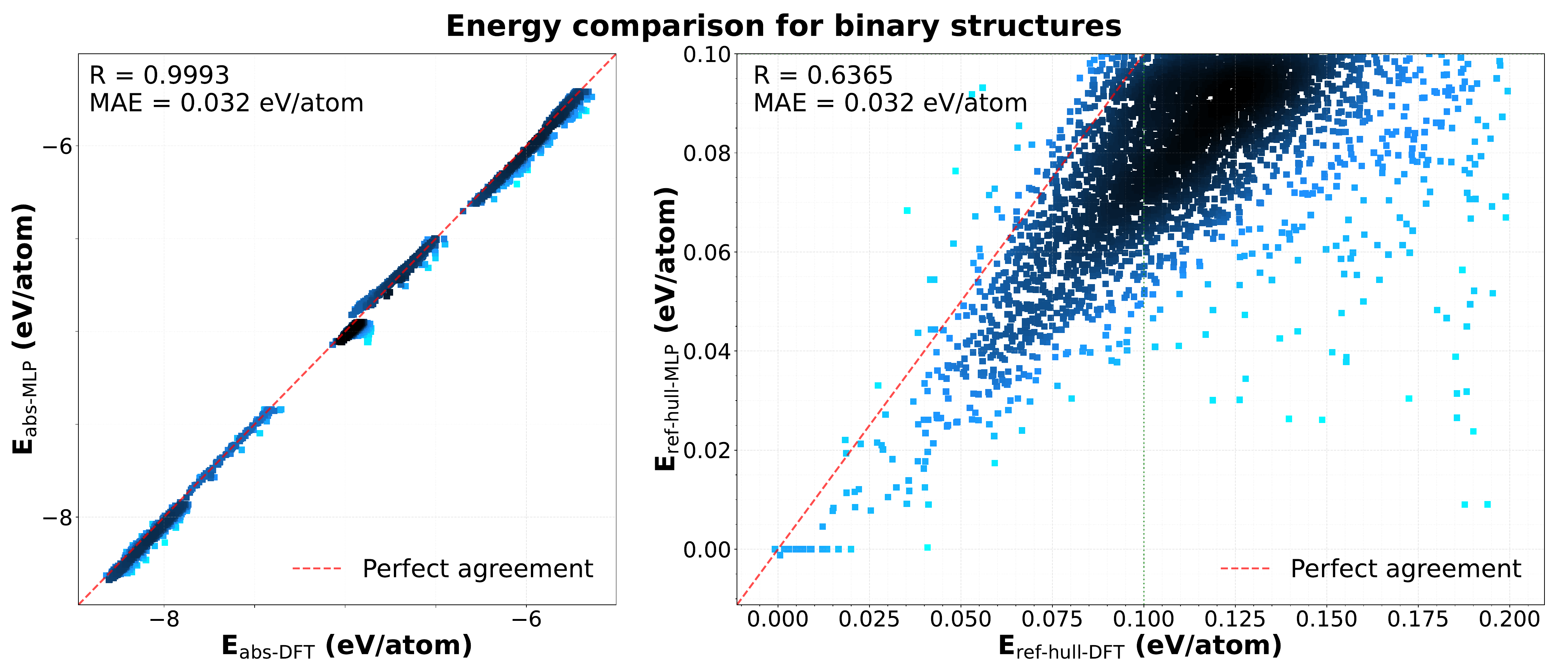}\\[1em]
  \includegraphics[width=0.95\textwidth]{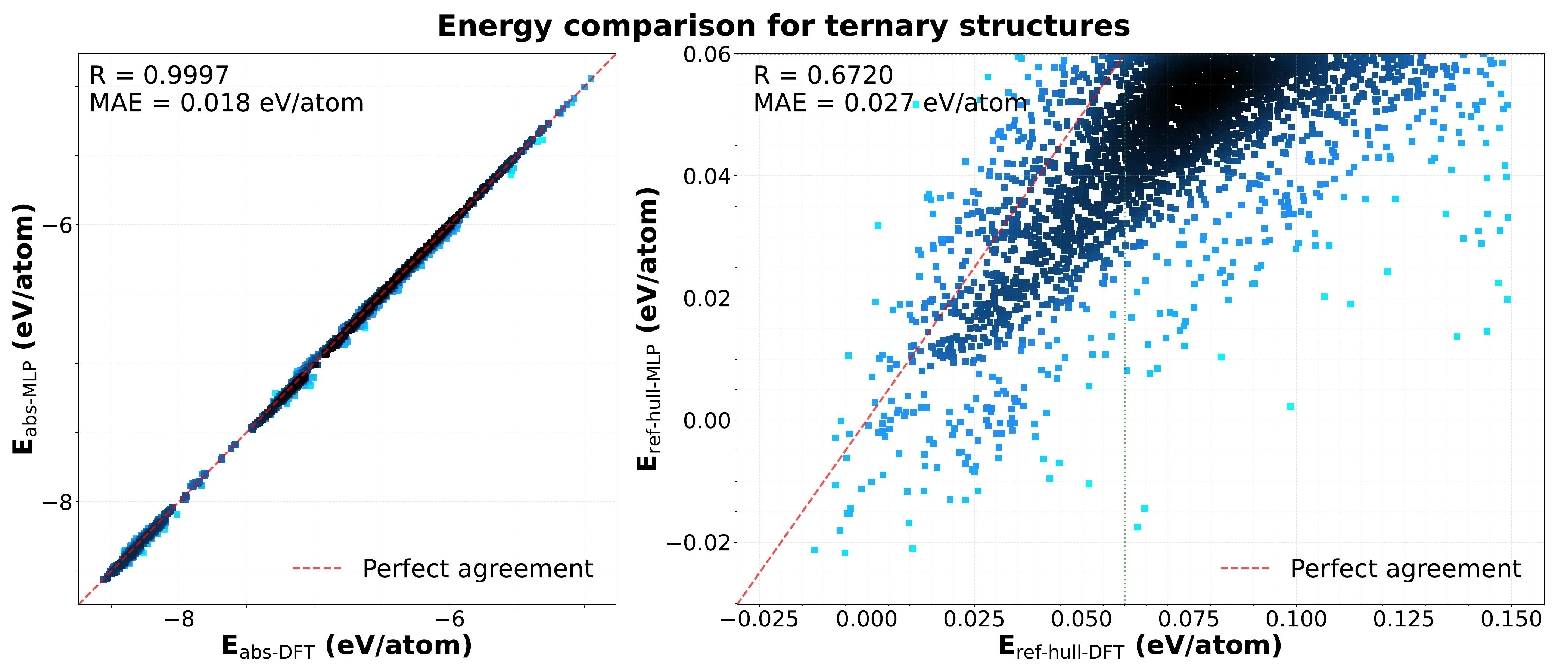}
  \caption{MLP vs.\ DFT energy comparison for binary (top) and ternary (bottom) structures in Y and Sm systems. Left panels: absolute energy per atom. Right panels: energy above hull referenced to MP phases.}
  \label{fig:scatter_bin_ter}
\end{figure*}

\subsection{Prescreening validation}\label{sec:prescreen_valid}

To assess prescreening reliability, we compared MLP and DFT energies
for all structures that passed prescreening and completed VASP relaxation
(see Fig.~\ref{fig:scatter_bin_ter} for full scatter plots).
Across both binary and ternary systems,
absolute energies per atom exhibit near-perfect MLP--DFT correlation
($R > 0.999$, MAE = \SIrange{0.013}{0.043}{eV/atom}),
confirming that the MLIP faithfully reproduces the DFT energy landscape.
The energy above hull (referenced to MP phases) shows weaker correlation
($R = 0.40$--$0.65$) because the hull distance amplifies
small absolute-energy errors through subtraction of
composition-dependent reference energies.
Despite this, the MLP prescreening retains the vast majority
of DFT-stable structures while reducing the VASP workload
by more than an order of magnitude (see Table~\ref{tab:funnel}).
All 5 structures that lie on the true DFT convex hull
passed the MLP prescreening, indicating that the approach does
not discard thermodynamically stable candidates.

\subsection{Summary of Low-energy Compounds}\label{sec:binary}

Across Y and Sm systems, 110 binary candidates satisfy
$\ehull^\mathrm{DFT\text{-}ref} < \SI{0.1}{eV/atom}$ with space-group number $\ge 75$
(60 for Y, 50 for Sm).
Of these, 78 ($\sim$71\%) are structurally novel (no match in the AFLOW
prototype library), while 32 match known binary prototypes.


\begin{table*}[htbp]
\caption{Selected binary R-T candidates.
Top section: ten lowest distance from the DFT hull.
Bottom section: ten highest magnetization.
$M$ is the total magnetic moment ($\mu_\mathrm{B}$);
$M/V$ is the magnetization ($\muBA$).
AFLOW\_prototype denotes the AFLOW prototype label
(Pearson symbol and space-group number appended).
\textbf{Bold rows} denote newly discovered candidates
which are not in the AFLOW prototype database.}
\label{tab:top_binary}
\begin{ruledtabular}
\begin{tabular}{lcrrrrl}
\multicolumn{1}{l}{\multirow{2}{*}{Structure}} & \multicolumn{1}{c}{$\ehull^\mathrm{DFT}$} & \multicolumn{1}{c}{$M$} & \multicolumn{1}{c}{$V$} & \multicolumn{1}{c}{$M/V$} & \multicolumn{1}{c}{$\mu_0 M/V$} & \multicolumn{1}{l}{\multirow{2}{*}{AFLOW\_prototype}} \\
 & \multicolumn{1}{c}{(eV/atom)} & \multicolumn{1}{c}{($\mu_\mathrm{B}$)} & \multicolumn{1}{c}{(\AA$^3$)} & \multicolumn{1}{c}{($\muBA$)} & \multicolumn{1}{c}{(T)} & \\
\midrule
\multicolumn{7}{l}{Top 10 Lowest DFT $\ehull$ structures} \\
YFe$_3$       & $0.000$ &   49.2 &  546.9 & 0.090 & 1.048 & A3B\_hR12\_166$^a$ \\
Y$_2$Fe$_7$       & $0.001$ &   77.9 &  802.4 & 0.097 & 1.131 & A7B2\_hR18\_166$^a$ \\
YCo$_3$       & $0.002$ &   30.7 &  523.9 & 0.059 & 0.682 & AB3\_hR12\_166$^a$ \\
Y$_2$Co$_{17}$    & $0.003$ &   74.9 &  720.3 & 0.104 & 1.212 & A17B2\_hR19\_166$^a$ \\
YNi$_3$       & $0.003$ &    0.0 &  522.2 & 0.000 & 0.000 & AB3\_hR12\_166$^a$ \\
Y$_2$Ni$_7$       & $0.005$ &    8.9 &  761.3 & 0.012 & 0.136 & A7B2\_hR18\_166$^a$ \\
Sm$_2$Co$_{17}$   & $0.005$ &   75.1 &  732.8 & 0.103 & 1.195 & A17B2\_hR19\_166$^a$ \\
SmCo$_3$      & $0.007$ &   30.8 &  538.1 & 0.057 & 0.667 & AB3\_hR12\_166$^a$ \\
Sm$_2$Ni$_7$      & $0.009$ &    7.8 &  778.1 & 0.010 & 0.116 & A7B2\_hR18\_166$^a$ \\
YNi$_5$       & $0.009$ &    1.7 &   81.5 & 0.021 & 0.246 & AB5\_hP6\_191$^a$ \\
\midrule
\multicolumn{7}{l}{Top 10 highest magnetization structures} \\
SmFe$_{12}$   & $0.027$ &   53.7 &  340.9 & 0.158 & 1.837 & A12B\_tI26\_139$^a$ \\
YFe$_{12}$    & $0.018$ &   51.6 &  334.7 & 0.154 & 1.797 & A12B\_tI26\_139$^a$ \\
Y$_2$Fe$_{17}$    & $0.023$ &  119.7 &  777.8 & 0.154 & 1.794 & A17B2\_hR19\_166$^a$ \\
Sm$_2$Fe$_{17}$   & $0.027$ &  119.9 &  788.5 & 0.152 & 1.771 & A17B2\_hR19\_166$^a$ \\
\textbf{SmFe$_{16}$}  & $\mathbf{0.042}$ & \textbf{64.5} & \textbf{427.3} & \textbf{0.151} & \textbf{1.759} & \textbf{AB16\_tI34\_139}$^c$ \\
\textbf{SmFe$_{16}$}  & $\mathbf{0.043}$ & \textbf{31.6} & \textbf{213.3} & \textbf{0.148} & \textbf{1.724} & \textbf{AB16\_tP17\_123}$^c$ \\
\textbf{SmFe$_{12}$}  & $\mathbf{0.058}$ & \textbf{23.3} & \textbf{166.9} & \textbf{0.139} & \textbf{1.625} & \textbf{AB12\_tP13\_123}$^c$ \\
\textbf{YFe$_{12}$}   & $\mathbf{0.055}$ & \textbf{23.0} & \textbf{166.1} & \textbf{0.139} & \textbf{1.615} & \textbf{AB12\_tP13\_123}$^c$ \\
\textbf{YFe$_5$}      & $\mathbf{0.089}$ & \textbf{33.6} & \textbf{256.8} & \textbf{0.131} & \textbf{1.526} & \textbf{AB5\_hR6\_160}$^c$ \\
\textbf{Y$_3$Fe$_{16}$}   & $\mathbf{0.065}$ & \textbf{105.3} & \textbf{806.8} & \textbf{0.131} & \textbf{1.521} & \textbf{A3B16\_hR19\_160}$^c$ \\
\end{tabular}
\end{ruledtabular}
$^a$Direct match.\quad $^c$New structure.
\end{table*}

Table~\ref{tab:top_binary} lists the candidates closest to the
DFT convex hull and those with the highest magnetization.
The highest magnetizations are found in Fe-rich systems:
SmFe$_{12}$ (1.837 T),
YFe$_{12}$ (1.797 T) and several others with the calculated DFT $\ehull$ values ranging from 0.018 to 0.089 eV/atom.
Likewise, the Co-rich systems show moderate magnetization with better thermodynamic
stability (e.g., Y$_2$Co$_{17}$ at DFT $\ehull = \SI{0.003}{eV/atom}$
with $M_s$=1.212 T).
And the Ni-rich binary candidates generally exhibit low total
magnetization, consistent with Ni's weak itinerant magnetism.


The ternary screening yielded 234 candidates
(125 for Y, 109 for Sm).
Among these, 139 ($\sim$59\%) are structurally novel,
91 ($\sim$39\%) are classified as ``binary derived''
(matching a known binary prototype after merging T and T$'$ sublattices),
and 4 ($\sim$2\%) match known ternary prototypes directly. Table~\ref{tab:top_ternary} summarizes the ternary candidates
closest to the DFT convex hull and those with
the highest magnetization.
The highest ternary magnetizations are found in
Fe-Ti and Fe-Mn substituted systems:
YFe$_{18}$Ti (1.799 T) and
Sm$_2$Fe$_{16}$Mn (1.733 T). Four ternary structures are calculated to be stable on the DFT calculated convex hull, including
Y$_2$Co$_{16}$Ti (spg.\ 160),
Y$_2$Co$_{16}$Mn (spg.\ 160),
YFe$_9$Ti$_2$ (spg.\ 166, direct ternary match),
and Sm$_2$Co$_{16}$Mn (spg.\ 160).
The most prominent parent prototype is Th$_2$Zn$_{17}$ type
(A17B2\_hR19\_166), with Ti or Mn substituting
onto one TM Wyckoff site.
Ti-doped Co-rich systems are particularly favorable thermodynamically,
appearing repeatedly among the most stable candidates.

\begin{table*}[htbp]
\caption{Selected ternary R-T-T$'$ candidates.
Format follows Table~\ref{tab:top_binary}.}
\label{tab:top_ternary}
\begin{ruledtabular}
\begin{tabular}{lcrrrrl}
\multicolumn{1}{l}{\multirow{2}{*}{Structure}} & \multicolumn{1}{c}{$\ehull^\mathrm{DFT}$} & \multicolumn{1}{c}{$M$} & \multicolumn{1}{c}{$V$} & \multicolumn{1}{c}{$M/V$} & \multicolumn{1}{c}{$\mu_0 M/V$} & \multicolumn{1}{l}{\multirow{2}{*}{AFLOW\_prototype}} \\
 & \multicolumn{1}{c}{(eV/atom)} & \multicolumn{1}{c}{($\mu_\mathrm{B}$)} & \multicolumn{1}{c}{(\AA$^3$)} & \multicolumn{1}{c}{($\muBA$)} & \multicolumn{1}{c}{(T)} & \\
\midrule
\multicolumn{7}{l}{Top 10 Lowest DFT $\ehull$ structures} \\
Y$_2$Co$_{16}$Mn      & $0.000$ &  79.5 & 726.1 & 0.109 & 1.276 & A17B2\_hR19\_166$^b$ \\
Y$_2$Co$_{16}$Ti      & $0.000$ &  65.3 & 732.4 & 0.089 & 1.038 & A17B2\_hR19\_166$^b$ \\
YFe$_9$Ti$_2$         & $0.000$ &  45.3 & 487.3 & 0.093 & 1.082 & AB2C9\_hR12\_166$^a$ \\
Sm$_2$Co$_{16}$Mn     & $0.000$ &  79.8 & 739.7 & 0.108 & 1.257 & A17B2\_hR19\_166$^b$ \\
SmFe$_9$Ti$_2$        & $0.002$ &  45.3 & 491.3 & 0.092 & 1.074 & AB2C9\_hR12\_166$^a$ \\
Sm$_2$Ni$_9$Cu        & $0.004$ &   2.4 & 166.3 & 0.015 & 0.170 & AB5\_hP6\_191$^b$ \\
Sm$_3$Ni$_{14}$Cu     & $0.005$ &   4.0 & 250.1 & 0.016 & 0.186 & AB5\_hP6\_191$^b$ \\
Sm$_2$Co$_{16}$Ti     & $0.006$ &  42.4 & 495.8 & 0.086 & 0.997 & A17B2\_hR19\_166$^b$ \\
Sm$_2$Co$_{15}$Mn$_2$     & $0.006$ &  85.6 & 748.1 & 0.114 & 1.334 & A17B2\_hR19\_166$^b$ \\
Sm$_3$Ni$_{14}$Cu     & $0.006$ &   4.1 & 249.8 & 0.016 & 0.192 & AB5\_hP6\_191$^b$ \\
\midrule
\multicolumn{7}{l}{Top 10 highest magnetization structures} \\
\textbf{YFe$_{18}$Ti}    & $\mathbf{0.063}$ & \textbf{118.1} & \textbf{765.0} & \textbf{0.154} & \textbf{1.799} & \textbf{ABC18\_hR20\_160}$^c$ \\
Sm$_2$Fe$_{16}$Mn     & $0.045$ & 116.7 & 784.6 & 0.149 & 1.733 & A17B2\_hR19\_166$^b$ \\
\textbf{YFe$_{18}$V}     & $\mathbf{0.063}$ & \textbf{111.8} & \textbf{752.8} & \textbf{0.148} & \textbf{1.731} & \textbf{ABC18\_hR20\_160}$^c$ \\
Y$_2$Fe$_{16}$Mn      & $0.041$ & 114.4 & 771.3 & 0.148 & 1.729 & A17B2\_hR19\_166$^b$ \\
Y$_2$Fe$_{15}$Mn$_2$      & $0.056$ & 106.4 & 762.0 & 0.140 & 1.627 & A17B2\_hR19\_166$^b$ \\
\textbf{YFe$_{12}$Ti}    & $\mathbf{0.085}$ &  \textbf{77.2} & \textbf{558.2} & \textbf{0.138} & \textbf{1.613} & \textbf{ABC12\_hR14\_166}$^c$ \\
\textbf{YFe$_{17}$Ti$_2$}    & $\mathbf{0.057}$ & \textbf{107.2} & \textbf{775.8} & \textbf{0.138} & \textbf{1.611} & \textbf{AB2C17\_hR20\_160}$^c$ \\
YFe$_{11}$V       & $0.095$ &  23.2 & 171.0 & 0.136 & 1.581 & A12B\_hP13\_191$^b$ \\
YCo$_{10}$Mn$_2$      & $0.086$ &  21.9 & 164.2 & 0.134 & 1.557 & A12B\_hP13\_191$^b$ \\
\textbf{YCo$_{18}$Mn}    & $\mathbf{0.066}$ &  \textbf{95.0} & \textbf{717.7} & \textbf{0.132} & \textbf{1.543} & \textbf{ABC18\_hR20\_160}$^c$ \\
\end{tabular}
\end{ruledtabular}
$^a$Direct match.\quad $^b$Binary derived.\quad $^c$New structure.
\end{table*}

\begin{figure*}[htbp]
\centering
\includegraphics[width=0.9\textwidth]{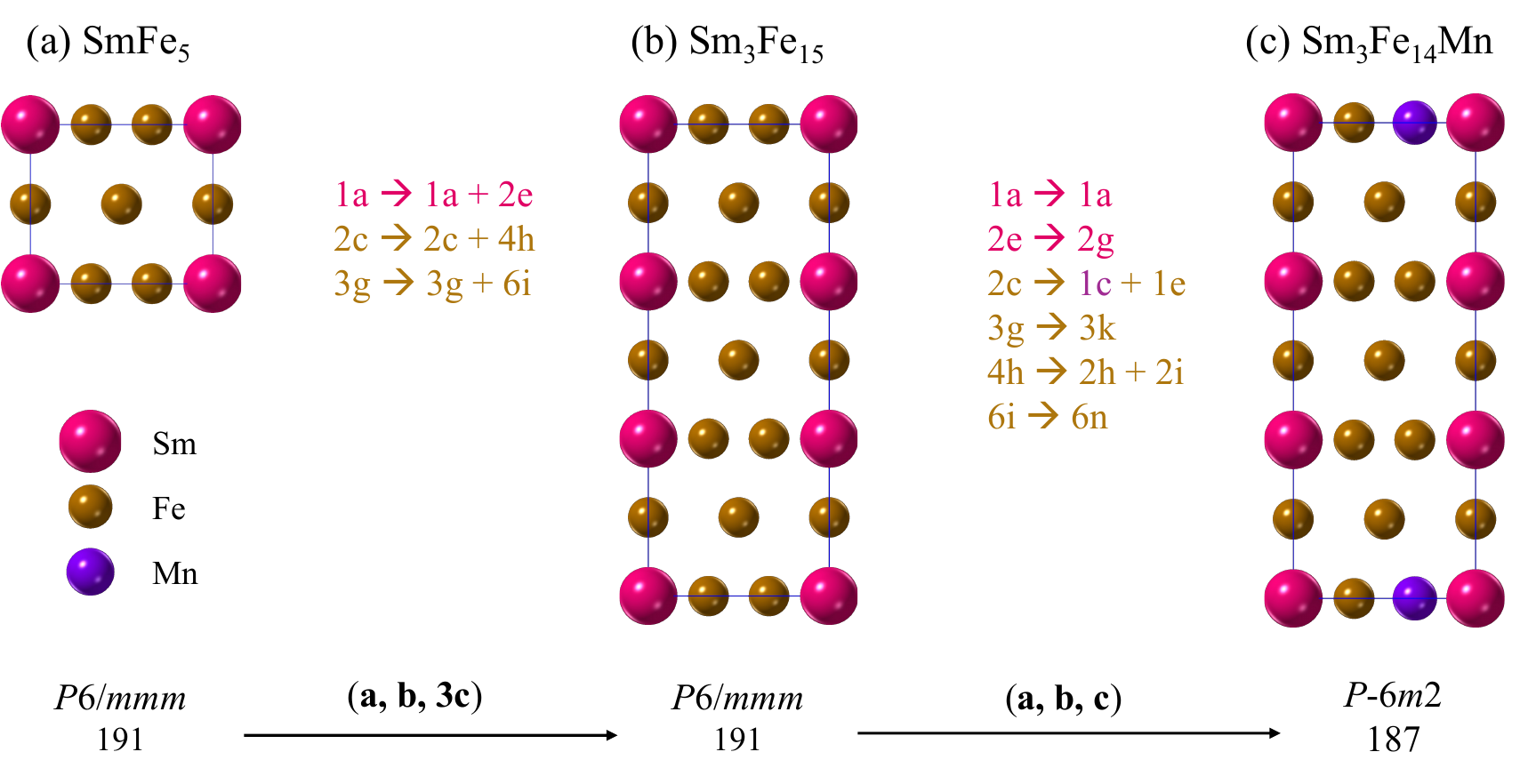}
\caption{A graphical illustration of binary-derived ternary compounds according to the symmetry relation from (a) original $P6/mmm$-SmFe$_5$, (b) intermediate $P6/mmm$-Sm$_3$Fe$_{15}$, to the ternary (c) $P\bar{6}m2$-Sm$_3$Fe$_{14}$Mn.}
\label{fig:sym}
\end{figure*}

\begin{figure*}[htbp]
\centering
\includegraphics[width=0.9\textwidth]{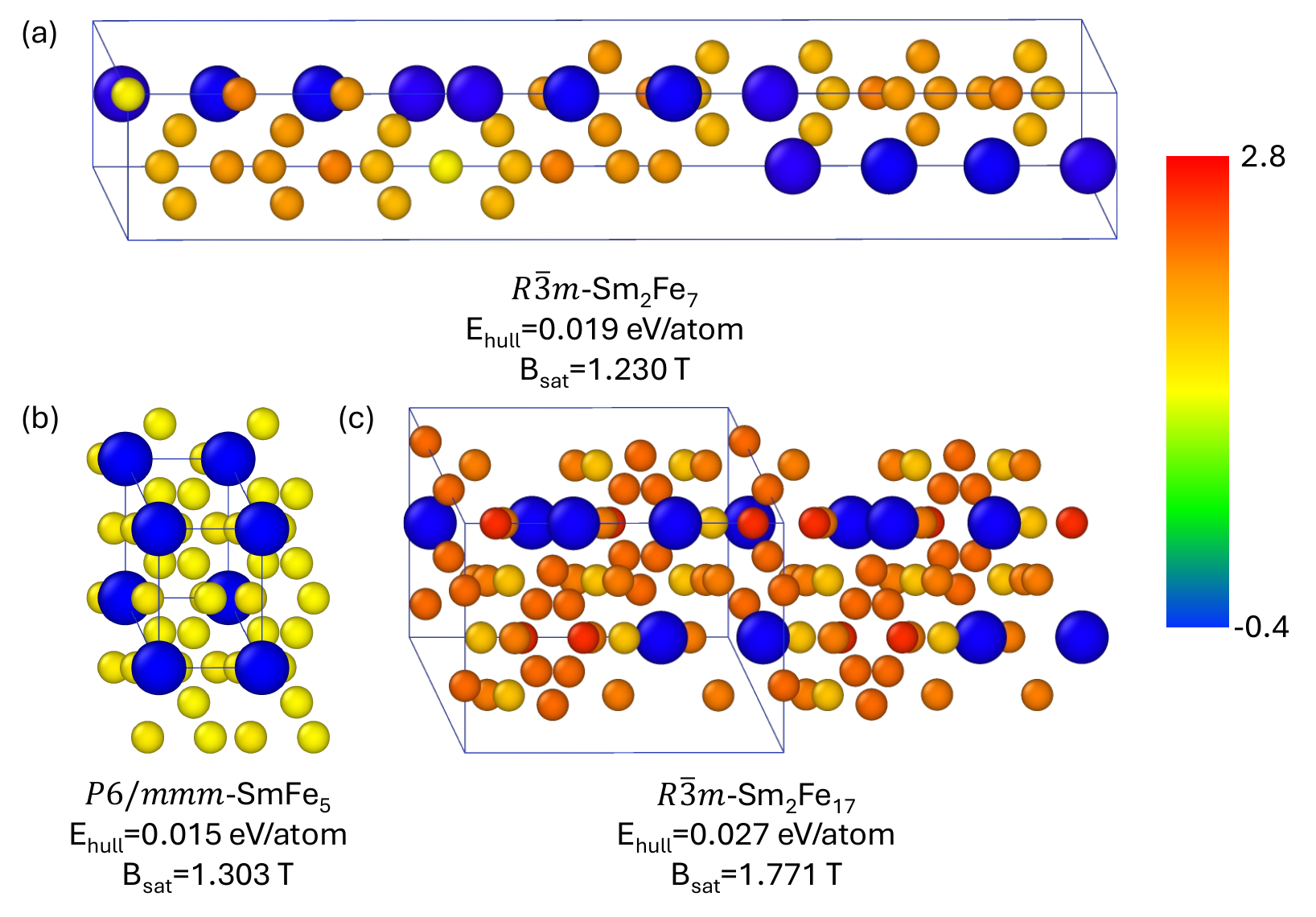}
\caption{DFT site-resolved magnetic moments for three Sm--Fe binary
candidates with increasing TM fraction:
(a)~Sm$_2$Fe$_7$,
(b)~SmFe$_5$,
and (c)~Sm$_2$Fe$_{17}$.
Atoms are colored by the on-site magnetic moment ($\mu_\mathrm{B}$/atom)
on a rainbow scale (blue $= -0.4$, red $= +2.8$);
sphere size distinguishes R (large) from TM (small) sublattices.
The saturation magnetization increases monotonically with the Fe fraction.}
\label{fig:binary_mag}
\end{figure*}

\subsection{Symmetry Relation Analysis}\label{sec:wyckoff}

By further looking into these structures, we found that a majority of the newly identified ternary candidates exhibit strong subgroup symmetry relationships with their binary parent compounds. 

Using our previously developed subgroup symmetry module within \texttt{PyXtal}~\cite{zhu2022symmetry}, we examine the SmFe$_5$-type parent phase (designated as AB5\_hP6\_191, space group 191, $P6/mmm$), in which RE atom occupies the $1a$ site and TM atoms fill two sublattices at $2c$ and $3g$. As shown in Fig.~\ref{fig:sym}a, this arrangement can be intuitively conceptualized as an alternating stacking of pure TM layers and RE-doped TM layers along the $c$-axis in a hexagonal lattice. Without altering the space group symmetry, this lattice can transition to an intermediate \textit{klassengleiche} subgroup representation by tripling the unit cell along the $c$-axis via Wyckoff site splitting ($1a \rightarrow 1a + 2e$; $2c \rightarrow 2c + 4h$; $3g \rightarrow 3g + 6i$; see Fig.~\ref{fig:sym}b). Furthermore, certain TM Wyckoff sites undergo additional splitting to accommodate the partial substitution of a secondary TM element (e.g., the $2c$-Fe site in $P6/mmm$ splits into $1c$-Mn + $1e$-Fe in $P\bar{6}m2$) following a \textit{translationengleiche} subgroup relationship, yielding a ternary system of Sm$_3$Fe$_{14}$Mn as depicted in Fig.~\ref{fig:sym}c.

Similar subgroup symmetry relationships are observed in the Th$_2$Zn$_{17}$ type compound, where TM atoms occupy four distinct sites: $9d$ ($\cdot 2/m$), $18f$ ($\cdot 2$), $18h$ ($\cdot m$), and $6c$ ($3m$). In this structure, two primary substitution scenarios emerge. First, early transition metals such as Ti, V, and Cr tend to fully occupy a specific Wyckoff site (e.g., $6c$) to preserve the high parent space group symmetry. Second, elements like Mn partially substitute across specific Wyckoff sites, lowering the symmetry to a subgroup representation. These site-substitution preferences are primarily driven by similarities in atomic radii and magnetic moments relative to Fe and Co. Overall, these substitution patterns provide systematic guidance for the targeted experimental synthesis of novel R--T permanent magnets.


\begin{figure*}[htbp]
\centering
\includegraphics[width=0.95\textwidth]{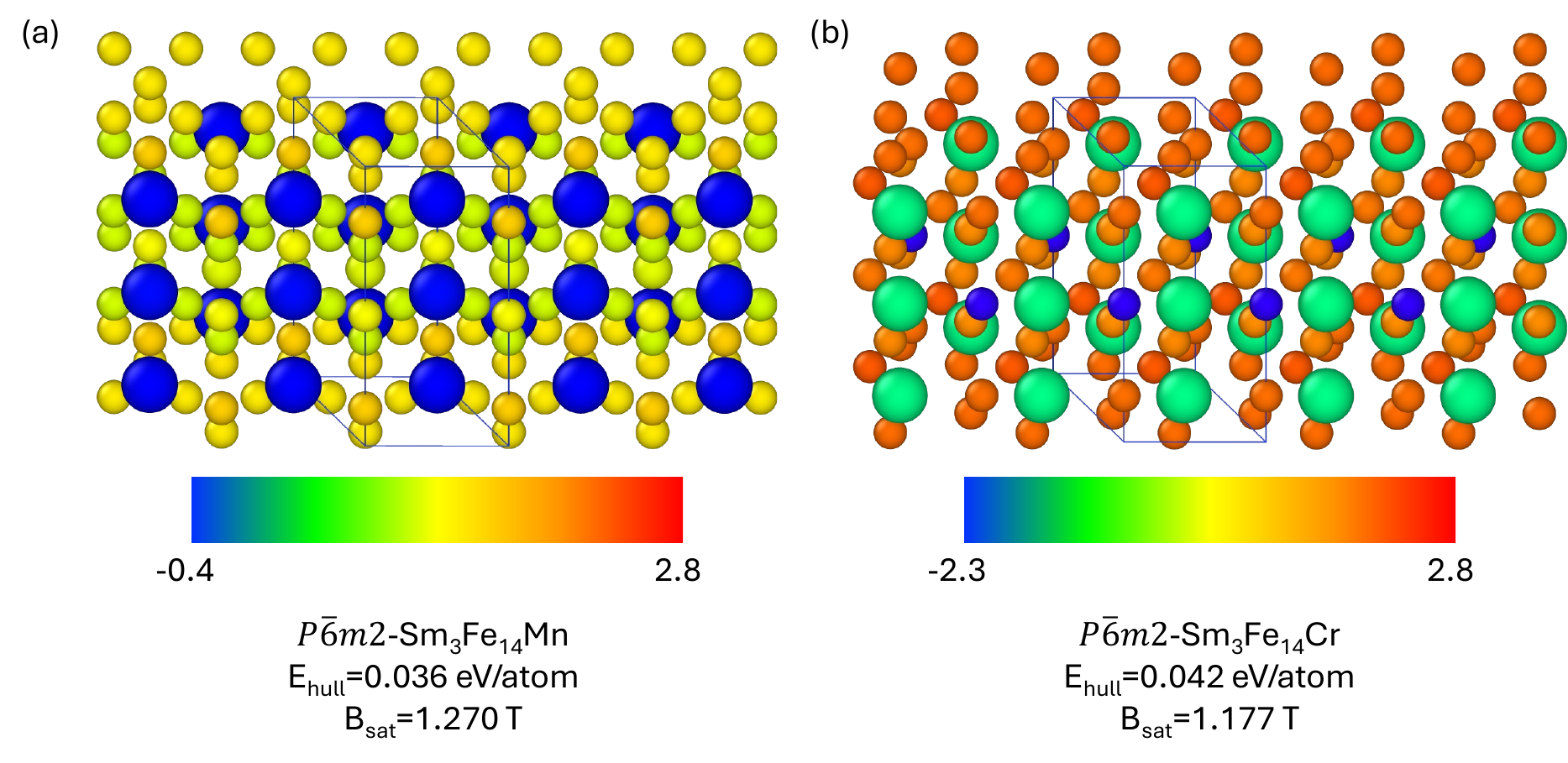}
\caption{DFT site-resolved magnetic moments for the CaCu$_5$-derived
ternary Sm--Fe systems:
(a)~Sm$_3$Fe$_{14}$Mn  and
(b)~Sm$_3$Fe$_{14}$Cr.
Atoms are colored by the on-site magnetic moment ($\mu_\mathrm{B}$/atom)
on a rainbow scale; note the different colorbar ranges
($-0.4$ to $+2.8$ for Mn-substituted vs.\ $-2.3$ to $+2.8$ for Cr-substituted)
to accommodate the strong antiferromagnetic Cr moment ($-2.29$~$\mu_\mathrm{B}$).
Mn aligns ferromagnetically with the Fe sublattice,
while Cr couples antiferromagnetically.}
\label{fig:ternary_tprime}
\end{figure*}

\subsection{Magnetic properties}\label{sec:mag}

The saturation magnetization $\mu_0 M/V$ serves as a first-order indicator
of the magnetic performance of each candidate.
Across all 344 candidates, the highest magnetizations
are concentrated in Fe-rich systems.
Binary Fe-rich phases reach saturation magnetization up to 1.837 T
(SmFe$_{12}$) and 1.797 T (YFe$_{12}$),
while ternary Fe-Ti substituted phases reach 1.799 T
(YFe$_{18}$Ti).
Co-rich systems show moderate magnetization
($\sim$1.257 T)
with better thermodynamic stability than their Fe-rich counterparts.
Ni-rich candidates generally exhibit weak magnetization ($\approx$ 0.583 T),
reflecting weak itinerant magnetism. Among the 5 stable candidates,
Y$_2$Co$_{16}$Mn combines thermodynamic stability with the highest magnetization among stable ternary phases
(1.276 T).

To understand the origin of these magnetization trends, we performed site-resolved magnetic moment analysis on representative Fe-rich binary candidates and their CaCu$_5$-derived ternary counterparts.

\paragraph{TM fraction and magnetization.}
Across the binary R-Fe series, the saturation magnetization increases monotonically with the TM fraction $R_\mathrm{TM} = N_\mathrm{TM}/(N_\mathrm{R}+N_\mathrm{TM})$.
For Y--Fe, the progression
YFe$_3$ ($R_\mathrm{TM}=0.75$, $\mu_0 M/V = 1.10$~T)
$\to$ Y$_2$Fe$_7$ ($0.78$, $1.18$~T)
$\to$ Y$_2$Fe$_{17}$ ($0.89$, $1.83$~T)
shows a nearly twofold increase in saturation magnetization;
an analogous trend holds for Sm--Fe
(Sm$_2$Fe$_7$ at $1.23$~T $\to$ SmFe$_5$ at $1.30$~T
$\to$ Sm$_2$Fe$_{17}$ at $1.77$~T),
as illustrated in Fig.~\ref{fig:binary_mag}.
Two factors drive this trend:
(i)~the R sublattice couples antiferromagnetically to TM
(average R moment $\approx -0.3$~$\mu_\mathrm{B}$),
so diluting R raises the net moment per volume;
(ii)~the average Fe moment itself increases from
${\sim}2.0$~$\mu_\mathrm{B}$ in 1:3 structures
to ${\sim}2.4$~$\mu_\mathrm{B}$ in 2:17 structures,
reflecting stronger exchange coupling in
Fe-dense coordination environments.

\paragraph{Effect of T$'$ substitution on magnetization.}
Fig.~\ref{fig:ternary_tprime}
compares the site-averaged magnetic moments
in the parent SmFe$_5$ with its CaCu$_5$-derived ternary derivatives
Sm$_3$Fe$_{14}$Mn and Sm$_3$Fe$_{14}$Cr.
Both substituents occupy the same multiplicity-1 Wyckoff site,
but their magnetic coupling differs markedly, as summarized in Table~\ref{tab:tprime}.
Mn couples ferromagnetically ($+1.96$~$\mu_\mathrm{B}$),
close to the Fe sublattice average,
resulting in minimal magnetization loss
($1.270$ vs.\ $1.303$~T in the parent).
Cr couples antiferromagnetically ($-2.29$~$\mu_\mathrm{B}$)
but enhances the moments of neighboring Fe atoms
to $+2.23$~$\mu_\mathrm{B}$ (vs.\ $+2.07$ in the parent),
partially compensating the cancellation
($1.177$~T).
These trends are consistent in the Y$_3$Fe$_{14}$T$'$ series,
confirming that the T$'$ substitution effect on magnetization
is governed by the interplay between dopant coupling polarity,
dopant moment magnitude, and the induced enhancement of the Fe sublattice.

\begin{table}[htbp]
\caption{Site-resolved magnetic moments and saturation magnetization
for the CaCu$_5$-derived binary parents and their T$'$-substituted
ternary derivatives.
$\bar{m}_\mathrm{R}$, $\bar{m}_\mathrm{Fe}$, $\bar{m}_{\mathrm{T}'}$:
average per-atom moment of each sublattice.}
\label{tab:tprime}
\begin{ruledtabular}
\begin{tabular}{lcrrrc}
Structure & T$'$ & \multicolumn{1}{c}{$\bar{m}_\mathrm{R}$} & \multicolumn{1}{c}{$\bar{m}_\mathrm{Fe}$} & \multicolumn{1}{c}{$\bar{m}_{\mathrm{T}'}$} & $\mu_0 M/V$ \\
 & & \multicolumn{1}{c}{($\mu_\mathrm{B}$)} & \multicolumn{1}{c}{($\mu_\mathrm{B}$)} & \multicolumn{1}{c}{($\mu_\mathrm{B}$)} & (T) \\
\midrule
\multicolumn{6}{l}{\textit{The Sm--Fe system}} \\
SmFe$_5$              &  & $-0.25$ & $+2.07$ &       & 1.303 \\
Sm$_3$Fe$_{14}$Mn     & Mn   & $-0.24$ & $+2.02$ & $+1.96$  & 1.270 \\
Sm$_3$Fe$_{14}$Cr     & Cr   & $-0.27$ & $+2.23$ & $-2.29$  & 1.177 \\
\midrule
\multicolumn{6}{l}{\textit{The Y--Fe system}} \\
Y$_2$Fe$_7$           &   & $-0.33$ & $+2.03$ &       & 1.178 \\
Y$_3$Fe$_{14}$Mn      & Mn   & $-0.25$ & $+2.00$ & $+1.90$  & 1.319 \\
Y$_3$Fe$_{14}$Cr      & Cr   & $-0.27$ & $+2.06$ & $-2.12$  & 1.164 \\
\end{tabular}
\end{ruledtabular}
\end{table}



\section{Discussion and Conclusion}\label{sec:conclusion}

In this work, we attempted to combine rapid AI generative model with hierarchical
screening to search for new stable magnets with high
saturation magnetization ($>$1.5 T).
Using our workflow, we systematically explored over 3000 binary R-T and ternary R-T-T$'$ compositions
with a total of 240{,}000 generated crystal structures through machine-learning interatomic potential prescreening and
spin-polarized DFT validation.
As a result, we have identified 300+ low-energy magnetic candidates (within 0.1~eV/atom above the convex hull) with the highest saturation magnetization reaching ${\sim}1.8$~T in Fe-rich systems. Overall, our approach demonstrates a generalizable strategy for targeted magnetic materials discovery across vast composition--structure spaces.

To maximize saturation magnetization,
3$d$ transition metals (Fe, Co, Ni) must align ferromagnetically,
as seen in Fe--Co binary alloys \cite{bardosMeanMagneticMoments1969, shamsutdinovMagnetizationStructureComposition2022}.
Although introducing RE elements and transition metal dopants
($\mathrm{T}'$) enhances magnetocrystalline anisotropy and
structural stability, their non-magnetic or antiparallel coupling
inevitably leads to a reduction in net magnetization.
In our current search, peak magnetization was realized in
$\text{SmFe}_{12}$ among binary systems and
$\text{YFe}_{18}\text{Ti}$ among ternaries,
suggesting that a high Fe/Co ratio is therefore critical for
achieving superior saturation magnetization.
For a ternary unit cell capped at 20 atoms,
$\text{YFe}_{18}\text{Ti}$ represents the practical upper limit
of the Fe fraction.
Given that the current MatterGen model is restricted to unit cells
consisting a maximum of 20 atoms, sampling even higher Fe ratios
was beyond the present search space.
Consequently, expanding continuous structural searches into larger
unit cells of more than 20 atoms using more powerful generative
models capable of traversing broader configurational spaces holds
strong promise for uncovering novel and stable phases with
saturation magnetizations exceeding 1.8~T.

Importantly, we emphasize two observations from this study. First, our symmetry analysis reveals that the majority of ternary compounds are subgroup derivatives resulting from Wyckoff site splitting of known binary structural prototypes such as (A$_2$B$_{17}$ and AB$_5$). This indicates that the current generative model may be primarily limited to generating new structures via the existing structural prototype via chemical substitution and symmetry breaking rather than discovering entirely novel prototypical motifs. 
Second, it is clear that high magnetization is strongly correlated with the Fe/Co ratio within the R-T-T$'$ system. Hence, a promising route for discovering high-magnetization stable compounds is to develop a more powerful generative models that can explore Fe-rich compositions in larger unit cells ($>$20 atoms) with enhanced structural novelty.
\vspace{3mm}

\begin{acknowledgments}
This research was sponsored by U.S. Department of Energy’s Advanced Research Projects Agency-Energy (ARPA-E) under the MAGNITO NOFO award number DE-AR0002158, as well as the Postdoctoral Hiring Program by Division of Research at UNC Charlotte.\end{acknowledgments}

\section*{Data availability}
The source code, instructions, as well as scripts used to calculate the results of this study, are available in \url{https://github.com/MaterSim/HT-Magflow}. The identified magnet structures are also interactively available via \url{https://mmi.charlotte.edu/magnets}.

\section*{Conflict of interest}
All authors declare that they have no conflict of interest.

\bibliography{main}

\end{document}